\documentclass[onecolumn,pra,aps,superscriptaddress,showpacs]{revtex4}
\usepackage{amsmath}
\usepackage{graphicx}

\begin{document}
\title{Exact solutions to the three-dimensional Gross-Pitaevskii equation with modulated radial nonlinearity}
\author{Wei-Ting Wang}
\author{Ying-Ying Li}
\affiliation{Department of Physics, Beijing Normal University, Beijing 100875, China}
\author{Shi-Jie Yang\footnote{Corresponding author: yangshijie@tsinghua.org.cn}}
\affiliation{Department of Physics, Beijing Normal University, Beijing 100875, China}
\affiliation{State Key Laboratory of Theoretical Physics, Institute of Theoretical Physics, Chinese Academy of Sciences£¬Beijing, 100190}
\begin{abstract}
We study the Bose-Einstein condensate trapped in a three-dimensional spherically symmetrical potential. Exact solutions to the stationary Gross-Pitaevskii equation are obtained for properly modulated radial nonlinearity. The solutions contain vortices with different winding numbers and exhibit the shell-soliton feature in the radial distributions.
\end{abstract}
\maketitle

The experimental realization of matter-wave solitons in weakly interacting atomic Bose-Einstein condensates (BECs) has attracted great attention in the past decade. Dark\cite{Denschlag}, bright\cite{Khaykovich}, and gap solitons\cite{Zobay}, as well as vortices\cite{Yang,Matthews} have been observed and widely studied. By applying the Feshbach resonance techniques, scientists can conveniently control the nonlinearities. Many nonlinear phenomena have been predicted by the manipulation of the scattering length either in time\cite{Malomed,Saito} or in space\cite{Rodas-Verde,Garc}. It has been used to generate bright solitons\cite{Partridge,Cornish} and induce collapse\cite{Donley}. In addition, bright solitons\cite{Ieda} and periodic wave solutions\cite{Yan} were obtained in spinor BECs. However, only one-dimensional matter-wave solitons have been created in experiments by using cigar-shaped traps to confine the condensate\cite{Strecker}. The realization of higher-dimensional matter-wave solitons is still a challengeable task because they are usually unstable for the nonlinear Schr\"{o}dinger (NLS) equation with constant or uniform couplings due to the weak and strong collapses of the BECs\cite{Sulem}. Modulation of atomic scattering length by the Feshbach resonance is expected to dynamically stabilize higher dimensional bright solitons\cite{Liu}.

On the other hand, a multi-quantized vortex is usually unstable when the BECs are trapped in harmonic potential\cite{Pu,Jackson,Virtanen,Isoshima,Nilsen}. It will split into several singly-quantized vortices. However, in the presence of a plug potential\cite{Simula}, an anharmonic trapping potential\cite{Lundh,Shaeer}, or an immiscible two-species BEC\cite{Yang,Wu}, the multi-quantized vortex can be effectively stabilized. For example, when the confining potential is steeper than the harmonic potential, multi-quantized vortices are energetically favorable.

In this paper, we study 3D solutions to the Gross-Pitaevskii equation (GPE) by means of the similar transformation. Exact solitary vortices in the BEC trapped in an external potential are constructed by properly modulating the radial nonlinearity. The number of vortex-soliton (VS) modes is specified by the radial nodes in the wavefunction and the winding number of the vortices.

The scaled stationary GPE for the macroscopic wave function reads
\begin{equation}
\mu\psi({\bf r})=-\nabla^2\psi({\bf r})+g({\bf r})|\psi({\bf r})|^2\psi({\bf r})+V({\bf r})\psi({\bf r}),
\end{equation}
where $V({\bf r})$ is the spherically harmonic oscillator potential. $g({\bf r})$ is the nonlinearity coefficient which is spatially variable. We rewrite the wavefunction in the spherical coordinates as
\begin{equation}
\psi(r,\theta,\varphi)=\Phi(r)Y_{lm}(\theta,\varphi),\label{wavefunction}
\end{equation}
where $Y_{lm}(\theta,\varphi)=P_l^{|m|}(\cos\theta)e^{im\varphi}$ ($l=0,1,2,\cdots$, and $m=0,\pm 1,\pm 2,\cdots,\pm l$) is the spherical harmonic function. The radial part $\Phi(r)$ obeys the following nonlinear equation,
\begin{equation}
\mu\Phi=-\Phi{''}-\frac{2}{r}\Phi{'}+g(r)\Phi^3+[\frac{l(l+1)}{r^2}+V(r)]\Phi.
\end{equation}
The convergence condition for $r\rightarrow 0$ requires that $\Phi(r)\sim r^{l}$ for $l\neq 0$ while $\Phi'(r)=0$ for $l=0$. In the meantime, $\Phi(r\rightarrow\infty)\rightarrow 0$ due to the localization of the BEC trapped in the potential.

According to the similar transformation, we define $\Phi(r)\equiv\rho(r)U[R(r)]$ so as to obtain two independent equations\cite{Liu},
\begin{equation}
 \rho{''}+\frac{2 \rho{'}}{r}+[-\frac{l(l+1)}{r^2}+\mu-V(r)]\rho=\frac{E}{r^4\rho^3},\label{eq1}
\end{equation}
and
\begin{equation}
  -\frac{d^2U}{dR^2}+g_0U^3=EU,\label{eq2}
\end{equation}
where $R(r)\equiv\int_0^r{x^{-2}\rho^{-2}(x)}dx$ which can be viewed as a rescaled radius. For this purpose, the spatial modulation of nonlinear coupling should be chosen as $g(r)\equiv g_0r^{-4}\rho^{-6}(r)$. Here $E$ and $g_0$ are constants. The nonlinearity-modulation is determined by $\rho(r)$. Equation (\ref{eq2}) is exactly solvable in terms of the Jacobian elliptic functions which gives rise to solitary form of wave. For Eq.(\ref{eq1}), physically meaningful solutions impose restrictions on $\rho(r)$. As $ r\rightarrow 0$, $\rho(r)$ behaves as $r^{-a}$ with $a\geq 2/3$. It should diverge for $r\rightarrow\infty$ so that the nonlinearity $g(r)$ is bounded and the integration in $R(r)$ converges.

We first consider the repulsive nonlinearity ($g_0>0$). The confining trap is chosen as $V(r)=\alpha r^2$. The existence of solutions for Eq.(\ref{eq2}) demands $E>0$. The VS solution is
\begin{equation}
U(R)=\sqrt{2(E-k^2)/g_0}\textrm{sn}(kR,p),
\end{equation}
where $k$ is the period of the Jacobian elliptical function $\textrm{sn}$. The modulus $p=\sqrt{E/k^2-1}$ imposes restriction on $k$ as $\sqrt{E/2}\leq k\leq\sqrt{E}$. To meet boundary conditions $\Phi(0)=\Phi(\infty)=0$, it requires $k=nK(\sqrt{E/k^2-1}/R(\infty))$ ($n=2,4,6,\cdots$). Here $K(p)$ is the complete elliptic integral of the first kind. It follows that $\sqrt{E/2}\leq nK(\sqrt{E/k^2-1}/R(\infty))\leq\sqrt{E}$. The integer number $n$ satisfies $n<n_{\textrm{max}}=2R(\infty)\sqrt{E}/\pi$, which implies that there is only a finite number of radial VS modes (or none, if $n_{\textrm{max}}<2$).

On the other hand, Eq.(\ref{eq1}) is a nonlinear equation which can be solved numerically. To perform the calculation, one requires $\rho \sim r^{-|s|}$ at $r\rightarrow 0$. Thus the nonlinear term in Eq.(\ref{eq1}) can be neglected near $r=0$. In this case, $\rho(r)$ behaves like the Neumann function $Y_s(\sqrt{\mu}r)$ at $r\rightarrow 0$ for $\mu>0$ (for $\mu<0$ it can be checked that the VS solutions do not exist). On the other end of $r\rightarrow\infty$, $\rho(r)\rightarrow\infty$ due to the presence of the trap. In addition, $Er^{-4}\rho^{-3}$ with $E>0$ on the righthand side of Eq.(\ref{eq1}) guarantees the sign definiteness of $\rho(r)$. Numerically, we take the initial function $\rho(r)$ by setting a small value of $r_0$ to the Neumann function $Y_s(\sqrt{\mu}r)$ and its derivative. The function $R(r)$ and $g(r)$ can be simultaneously obtained by adjusting the value of $s$ according to the angular momentum $l$.

\begin{figure}[h]
\begin{center}
\includegraphics[width=8cm]{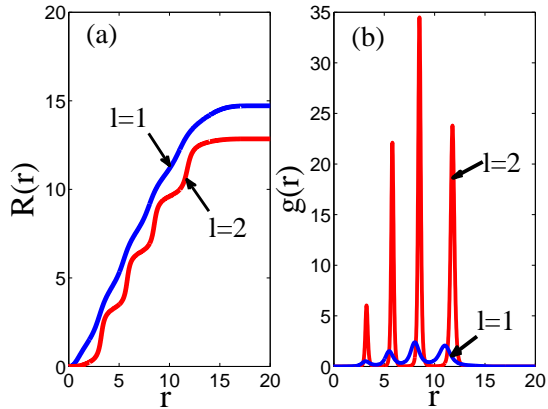}
\caption{(a) The rescaled radius $R(r)$ for $l=1$ (blue line) and $l=2$ (red line), respectively. (b) Modulation profiles of the repulsive nonlinearity $g(r)$ for $l=1$ (blue line) and $l=2$ (red line), respectively. The parameters are $E=1$, $\mu=2$, $\alpha=0.01$, and $g_0=0.01$.}
\end{center}
\end{figure}

\begin{figure}[h]
\begin{center}
\includegraphics[width=8cm]{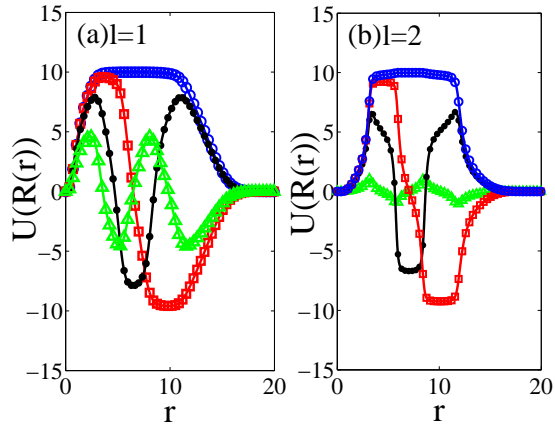}
\caption{$U(R)$ versus $r$ for (a) $l=1$, and (b) $l=2$. Open circles (blue), squares (red), solid circles (black), and triangles (green) denote solutions with $n = 2,4,6,8$, respectively. The parameters are the same as those in Fig.1. The number of nodes equals to $n/2-1$.}
\end{center}
\end{figure}

Figure 1 displays the numerical results of the rescaled radius $R(r)$ (Fig.1(a)) and the modulated nonlinearity $g(r)$ (Fig.1(b)) for $E=1$, $\mu=2$, $\alpha=0.01$, and $g_0=0.01$. We have chosen $s=1.38$ for $l=1$ and $s=2.56$ for $l=2$, respectively. It can be seen from Fig.1(a) that $R(r)$ is angular momentum-dependent and saturates as $r\rightarrow\infty$.  In Fig.1(b), the nonlinear modulation exhibits oscillations versus the radius. It vanishes at both $r=0$ and large distances. In addition, $g(r)$ is dependent on the angular momentum $l$. The radial oscillation of the nonlinear coupling $g(r)$ is related to the behavior of the function $\rho(r)$, which is an oscillating Whittaker Function as $E=0$ (see below). The oscillating feature is kept as $E\neq 0$. Figure 2 shows the analytical results for $U(r)$. Obviously, it satisfies the restrictions $U(R(0))=U(R(\infty))=0$ for both $l=1$ (Fig2.(a)) and $l=2$ (Fig.2(b)). The number of nodes increases as the radial quantum number $n$ increases.

\begin{figure}[h]
\begin{center}
\includegraphics[width=8cm]{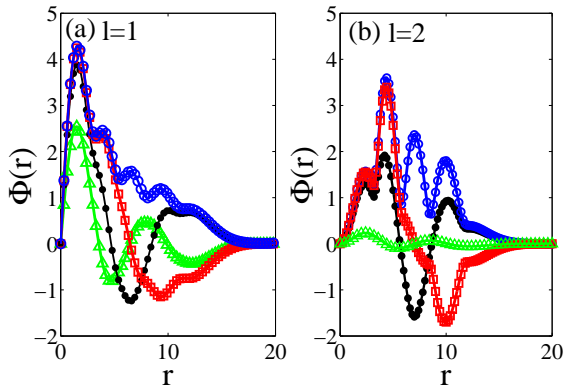}
\caption{The radial function $\Phi(r)$ for (a) $l=1$ and (b) $l=2$ in the presence of the external trap.
Open circles (blue), squares (red), solid circles (black), and triangles (green) denote solutions with $n = 2, 4, 6, 8$, respectively. The parameters are the same as those in Fig.1.}
\end{center}
\end{figure}

The radial wavefunctions $\Phi(r)$ are plotted in Fig.3 for $l=1$ (Fig.3(a)) and $l=2$ (Fig.3(b)), respectively, where $n=2,4,6,8$. Obviously, they are localized in the trap. We observe that the number of nodes is exactly equal to that of $U(r)$ as shown in Fig.2. It implies that the function $\rho(r)$ does not contribute nodes to the radial wavefunction. The whole 3D wavefunction is the combination of the radial and the angular part of the wavefunction, $\psi_{nlm}(r,\theta,\varphi)=\Phi_{nl}(r)Y_{lm}(\theta,\varphi)$. It is specified by three quantum numbers ($n/2,l,m$). The 3D density distributions are shown in Fig.4, where the upper row and lower row are for the radial quantum numbers $n=2$ and $n=6$, respectively. Fig.4(a) and (b) are for the angular momentum $l=1$ ($m=0,1$, respectively). Fig.4(c)-(e) are for $l=2$ ($m=0,1,2$, respectively).

\begin{figure}[h]
\begin{center}
\includegraphics[width=12cm]{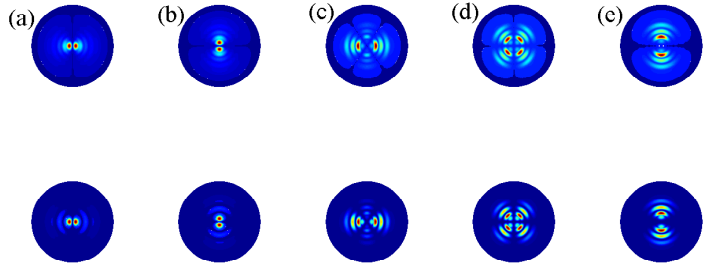}
\caption{The 3D density distributions of the states (\ref{wavefunction}) for $n=2$ (upper raw) and $n=6$ (lower raw). (a) $l=1,m=0$; (b) $l=1,m=1$; (c) $l=2,m=0$; (d) $l=2,m=1$; (e) $l=2,m=2$. The parameters are the same as those in Fig.1.}
\end{center}
\end{figure}

For the attractive nonlinearity ($g_0<0$), the results are similar to those of the repulsive nonlinearity as $E\neq 0$. Here we pay our attention to the case of $E=0$. Since the nonlinear term disappears, Eq.(\ref{eq1}) is exactly solvable for the harmonic trap $V(r)=\alpha r^2$. $\rho(r)$ can be represented in terms of the Whittaker' $M$ and $W$ Functions as $\rho(r)=r^{-3/2}[c_1M(\mu/4\sqrt{\alpha},(|l+1/2|)/2,\sqrt{\alpha}r^2)+c_2W(\mu/4\sqrt{\alpha},(|l+1/2|)/2,\sqrt{\alpha}r^2)]$. In the absence of the external potential, $\rho(r)$ degenerates into $\rho(r)=c_3j_{l}(\sqrt{-\mu}r)+c_4n_{l}(\sqrt{-\mu}r)$ with $\mu<0$, where $j_{l}$ and $n_{l}$ are the spherical Bessel and spherical Neuman functions of imaginary argument. Meanwhile, Eq.(\ref{eq2}) for $E=0$ yields to
\begin{equation}
U(R)=(n\eta/\sqrt{-g_0})\textrm{cn}(n\eta R-K(1/\sqrt{2}),1/\sqrt{2}),
\end{equation}
where the radial quantum number $n=2,4,6,\cdots$ and $\eta= K(\sqrt{2}/2)/R(\infty)$.

The results are displayed in Fig.5 for a harmonic trap. The relevant parameters are chosen as $\mu=-1$, $g_0 =-1$, and $\alpha=0.01$. Analogously, the number of nodes in the radial wavefunctions is determined by the number of nodes in $U(r)$.
\begin{figure}[h]
\begin{center}
\includegraphics[width=8cm]{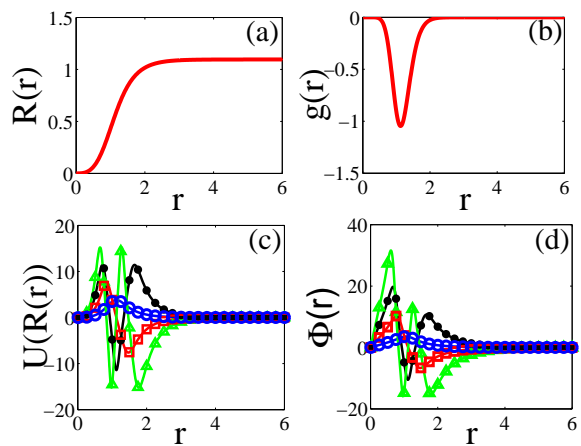}
\caption{Results for the attractive nonlinearity with $E=0$ and $l=1$. (a) The rescaled radius $R(r)$. (b) The modulated nonlinearity $g(r)$. (c) $U(r)$ versus $r$. (d) The radial distribution of the wavefunction. Open circles (blue), squares (red), solid circles (black), and triangles (green) denote solutions with $n = 2, 4, 6, 8$, respectively.}
\end{center}
\end{figure}

In summary, we have studied exact solitary solutions to the spherical symmetric GPE with radially modulated nonlinearity. The wavefunction can be expressed by the combination of a radial function and a spherical harmonic function. The number of VSs is limited and the quantum numbers can only be even numbers.

This work is supported by the funds from the Ministry of Science and Technology of China under Grant No. 2012CB821403.

\end{document}